\documentclass[12pt]{article}
\begin{document}
\begin{center}
{\bf The Holographic Principle and Dark Energy.
\\I.Deformed Quantum Field Theory and New Small Parameters}\\ \vspace{5mm}
A.E.Shalyt-Margolin \footnote{E-mail: alexm@hep.by;
a.shalyt@mail.ru}\\ \vspace{5mm} \textit{National Center of
Particles and High Energy Physics, Bogdanovich Str. 153, Minsk
220040, Belarus}
\end{center}
\vspace{0.5cm}
\begin{abstract}
In this work the Vacuum Energy Density Problem or Dark Energy
Problem is studied on the basis of the earlier results by the
author within the scope of the Holographic Principle.   It is
demonstrated that the previously introduced deformed quantum field
theory at a nonuniform lattice in the finite-dimensional hypercube
is consistent with the Holographic Principle (Holographic Entropy
Bound) in case the condition of the physical system's stability
with respect to the gravitational collapse is met, or simply
stated, the gravitational stability is constrained. The associated
deformation parameter is the basic characteristic, in terms of
which one can explain the essence of such a quantity as the vacuum
energy density and its smallness. Moreover, the entropy
characteristics are also well explained in terms of the above
deformation parameter.  The relation of this work to other studies
devoted to the Dark Energy Problem is considered. Besides, the
principal problems (tasks) are formulated; both the well-known
problems and those naturally following from the obtained results
\end{abstract}
\section{Introduction}
Recently the Vacuum Energy Density Problem or Dark Energy Problem
has become one of the key physical problems in basic research.
Numerous works and review papers on this problem have been
published in the last 10-15 years \cite{Dar1}. And a great number
of various approaches to this problem have been proposed: scalar
field models (quintessence model, K-essence, tachyon field,
phantom field, dilatonic, Chaplygin gas) \cite{Quint1},
\cite{K1},\cite{Tach1},\cite{Phant1},\cite{Dil1}, \cite{Chapl1},
braneworld models \cite{Brane1},dynamic approaches to the
cosmological constant $\Lambda$ \cite{Dyn1}, anthropic selection
of $\Lambda$ \cite{Antr1}, etc. By the author's opinion, however,
most promising are the approaches based on the Holographic
Principle \cite{Hooft1},\cite{Hooft2}. These approaches can
provide clear explanation \cite{Bal} for the theoretical smallness
of the vacuum energy density, and for the fact that this value
derived from astrophysical data is smaller by a factor of àbout
$10^{123}$ than that expected on the basis of a naive quantum
field theory \cite{Wein1}. Another approach known as the
Holographic Dark Energy also looks very promising (e.g.,
\cite{Hol1},\cite{Hol2}).
\\ The present work just deals with the vacuum energy
problem in the context of the Holographic Principle.
But unlike the majority of the papers devoted to the
subject, in this work a new approach to the problem is
developed by the author on the basis of deformation of
a quantum field theory, as an extension of his earlier
studies \cite{shalyt1}--\cite{shalyt10}. This work
involves two principal findings:
\\(1)Previously introduced by the author in
\cite{shalyt6},\cite{shalyt7},\cite{shalyt9}, a quantum field
theory at a nonuniform lattice $Lat_{\widetilde{\alpha}}$
dependent on the deformation parameter $\alpha$ , in case a
physical system is constrained gravitationally, meets the
Holographic Principle (or compatible with Holographic Principle);
\\(2) the deformation parameter $\alpha$ is the basic
characteristic, in terms of which one can clearly explain the
essence of  the vacuum energy density as well as its smallness.
Moreover, the associated entropy characteristics are also well
explained in terms of the above-mentioned parameter.
\\ The structure of the work is as follows. Section 2
presents short description of the principal results obtained
\cite{shalyt1}--\cite{shalyt10} on the density matrix deformation
in a Quantum Physics of the Early Universe (Plank Scale) and
associated with a new small parameter (deformation parameter)
$\alpha$ varying over the interval $(0,1/4]$. In Section 3 it is
demonstrated that the introduced quantum field theory at the
nonuniform lattice $Lat_{\widetilde{\alpha}}$ "existing" in the
finite-dimensional cube conforms well to the holographic principle
when the total energy of the Fock states is constrained
gravitationally. In Section 4 the main points of a holographic
approach to the vacuum energy density $\rho_{vac}$ in terms of
$\alpha$ are given, and it is demonstrated that smallness of
$\alpha$ suggests the smallness of $\rho_{vac}$. In Section 5 the
findings of this work are compared to the results of other authors
engaged in studies of the Vacuum Energy Density Problem. And the
last Section presents statement of the principal problems
including the well-known ones and direct inferences form the
author's results.
\section{Deformed Quantum Mechanics and Deformed
Quantum Field Theory} As it has been repeatedly demonstrated
earlier, a Quantum Mechanics of the Early Universe (Plank Scale)
is a Quantum Mechanics with the Fundamental Length
(QMFL)\cite{Gar1}--\cite{Magg1}. In the works by the author
\cite{shalyt1}--\cite{shalyt10} an approach to the construction of
QMFL has been developed with the help of the deformed density
matrix, the density matrix deformation in QMFL being a starting
object called the density pro-matrix and deformation parameter
(additional parameter) $\alpha=l_{min}^{2}/x^{2}$, where $x$ is
the measuring scale and $l_{min}\sim l_{p}$.
\\Exact definition will be as follows:
\cite{shalyt1},\cite{shalyt2},\cite{shalyt9}:
\\
\noindent {\bf Definition 1.} {\bf(Quantum Mechanics with the
Fundamental Length [for Neumann's picture])}
\\
\\
\noindent Any system in QMFL is described by a density pro-
matrix
of the form $${\bf
\rho(\alpha)=\sum_{i}\omega_{i}(\alpha)|i><i|
},$$ where \begin{enumerate}
\item Vectors $|i>$ form a full orthonormal
system; \item $\omega_{i}(\alpha)\geq 0$ and
for all $i$  the finite limit
$\lim\limits_{\alpha\rightarrow
0}\omega_{i}(\alpha)=\omega_{i}$ exists;
\item
$Sp[\rho(\alpha)]=\sum_{i}\omega_{i}(\alpha)<
1$, $\sum_{i}\omega_{i}=1.$;
\item For every operator $B$ and any $\alpha$
there is a mean operator $B$ depending on
$\alpha$:\\
$$<B>_{\alpha}=\sum_{i}\omega_{i}(\alpha)<i|B
|i>.$$ \item The following condition should
be fulfilled:
\begin{equation}\label{U1}
Sp[\rho(\alpha)]-
Sp^{2}[\rho(\alpha)]\approx\alpha.
\end{equation}
Consequently, we can find the value for $Sp[\rho(\alpha)]$
satisfying the above-stated condition:
\begin{equation}\label{U2}
Sp[\rho(\alpha)]\approx\frac{1}{2}+\sqrt{\frac{1}{
4}-\alpha} \end{equation}
and therefore
\item $0<\alpha\leq1/4$.
\end{enumerate}
It is no use to enumerate all the evident implications and
applications of {\bf Definition 1}, better refer to
\cite{shalyt1},\cite{shalyt2}. Nevertheless, it is clear that \\
{\bf for $\alpha\rightarrow 0$ the above limit covers both the
Classical and Quantum Mechanics depending on $\hbar\rightarrow 0$
or not}.
\\It should be noted that according to
{\bf Definition 1}, a minimum measurable length is equal
 to $l^{*}_{min}=2l_{min}$ being a nonreal
number at point $l_{min}$,$Sp[\rho(\alpha)]$. Because of this, a
space part of the Universe is a lattice with a spacing of
$a_{min}=2l_{min}\sim 2l_{p}$. In consequence the first issue
concerns the lattice spacing of any lattice-type model: a selected
lattice spacing $a_{lat}$ should not be less than $a_{min}$,i.e.
always
\\
$$a_{lat}\geq a_{min}>0$$
\\
Besides, a continuum limit in any lattice-type model is meaning
$a_{lat}\rightarrow a_{min}>0$ rather than $a_{lat}\rightarrow 0$.
\\ Proceeding from $\alpha$, for each space dimension we have a
discrete series of rational values for the inverse squares of even
numbers nonuniformly distributed along the real number line
$\alpha = 1/4, 1/16,1/36,1/64,...$. A problem arises, is this
series somewhere terminated or, on the contrary, is it infinite?
\\Provided, the series is finite, we have
\\
$$0<l^{2}_{min}/l^{2}_{max}\leq\alpha\leq1/4.$$
\\
\\In Sections 4 and 5 of this work $l_{max}$
is naturally following from the problem of
the nonzero cosmological constant $\Lambda$.
\\Note that in the majority of cases all
three space dimensions are equal, at least at large scales, and
hence their associated values of $\alpha$ parameter should be
identical. This means that for most cases, at any rate in the
large-scale (low-energy) limit, a single deformation parameter
$\alpha$ is sufficient to accept one and the same value for all
three dimensions to a high degree of accuracy. In the general
case, however, this is not true, at least for very high energies
(on the order of the Planck's), i.e. at Planck scales, due to
noncommutativity of the spatial coordinates
\cite{Ahl1},\cite{Magg1}:
\\
$$\left[ x_i ,x_j \right]\neq 0$$
\\
As a result, in the general case we have a point with coordinates
${\bf \widetilde{\alpha}}=(\alpha_{1},\alpha_{2},\alpha_{3})$ in
the normal(three-dimensional) cube $I_{1/4}^{3}$ of side
$I_{1/4}=(0;1/4]$.
\\It should be noted that this universal cube may be extended to the
four-dimensional hypercube by inclusion of the additional parameter
$\tau,\tau\in I_{1/4}$ that is generated by internal energy of the
statistical ensemble and its temperature for the events when this
notion is the case. It will be recalled that $\tau$ parameter occurs
from a maximum temperature that is  in its turn generated by the
Generalized Uncertainty Relations of "energy - time" pair in GUR
\cite{shalyt3},\cite{shalyt9}.In both cases the generated series has
one and the same discrete set of values of parameter $\tau$ :$\tau =
1/4,1/16,1/36,1/64,...,1/4n^{2}... . $
\\Using $Lat_{\widetilde{\alpha}}$ we denote the lattice
in cube $I_{1/4}^{3}$ formed by points $\widetilde{\alpha}$, and
through $Lat^{\tau}_{\widetilde{\alpha}}$ we denote the lattice in
hypercube $I_{1/4}^{4}$, that is formed by points
$\widetilde{\alpha}_{\tau}=(\widetilde{\alpha},\tau)$.
\\Any quantum theory may be defined for the indicated lattice in
hypercube. To this end it is required to go from Neumann's picture to
Shr{\"o}dinger's picture. We recall the fundamental definition
\cite{shalyt2},\cite{shalyt6},\cite{shalyt9} with $\alpha$ changed by
$\widetilde{\alpha}$:
\\ \noindent {\bf Definition
$1^{\prime}$}{\bf QMFL (Shr{\"o}dinger's picture)}
\\
Here, the prototype of Quantum Mechanical normed wave function (or
the pure state prototype) $\psi(q)$ with $\int|\psi(q)|^{2}dq=1$
in QMFL is
$\psi(\widetilde{\alpha},q)=\theta(\widetilde{\alpha})\psi(q)$.
The parameter of deformation $\widetilde{\alpha}\in I_{1/4}^{3}$.
Its properties are
$|\theta(\widetilde{\alpha})|^{2}<1$,$\lim\limits_{|\widetilde{\alpha}|
\rightarrow 0}|\theta(\widetilde{\alpha})|^{2}=1$ and the relation
$|\theta(\alpha_{i})|^{2}-|\theta(\alpha_{i})|^{4}\approx
\alpha_{i}$ takes place. In such a way the total probability
always is less than 1:
$p(\widetilde{\alpha})=|\theta(\widetilde{\alpha})|^{2}
=\int|\theta(\widetilde{\alpha})|^{2}|\psi(q)|^{2}dq<1$ tending to
1, when  $\|\widetilde{\alpha}\|\rightarrow 0$. In the most
general case of the arbitrarily normed state in QMFL(mixed state
prototype)
$\psi=\psi(\widetilde{\alpha},q)=\sum_{n}a_{n}\theta_{n}(\widetilde{\alpha})\
psi_{n}(q)$ with $\sum_{n}|a_{n}|^{2}=1$ the total probability is
$p(\widetilde{\alpha})=\sum_{n}|a_{n}|^{2}|\theta_{n}(\widetilde{\alpha})|^{2
}<1$ and
 $\lim\limits_{\|\widetilde{\alpha}\|\rightarrow 0}p(\widetilde{\alpha})=1$.
It is natural that Shr{\"o}dinger equation is also deformed in
QMFL. It is replaced by the equation
\begin{equation}\label{U24S}
\frac{\partial\psi(\widetilde{\alpha},q)}{\partial t}
=\frac{\partial[\theta(\widetilde{\alpha})\psi(q)]}{\partial
t}=\frac{\partial\theta(\widetilde{\alpha})}{\partial
t}\psi(q)+\theta(\widetilde{\alpha})\frac{\partial\psi(q
)}{\partial t},
\end{equation}
where the second term in the right-hand side generates
the Shr{\"o}dinger equation as
\begin{equation}\label{U25S}
\theta(\widetilde{\alpha})\frac{\partial\psi(q)}{\partial
t}=\frac{-i\theta(\widetilde{\alpha})}{\hbar}H\psi(q).
\end{equation}
Here $H$ is the Hamiltonian and the first member is added
similarly to the member that appears in the deformed Liouville
equation, vanishing when $\theta[\widetilde{\alpha}(t)]\approx
const$. In particular, this takes place in the low energy limit in
QM, when $\|\widetilde{\alpha}\|\rightarrow 0$. It should be noted
that the above theory is not a time reversal of QM because the
combination $\theta(\widetilde{\alpha})\psi(q)$ breaks down this
property in the deformed Shr{\"o}dinger equation. Time reversal is
conserved only in the low energy limit, when a quantum mechanical
Shr{\"o}dinger equation is valid.
\\ According to {\bf Definition $1^{\prime}$}, everywhere
$q$ is the coordinate of a point at the three-dimensional space.
As indicated in \cite{shalyt1}--\cite{shalyt10}, for a density
pro-matrix there exists an exponential ansatz satisfying the
formula \ref{U1} in {\bf Definition 1}:
\begin{equation}\label{U26S}
\rho^{*}(\alpha)=\sum_{i}\omega_{i} exp(-\alpha)|i><i|,
\end{equation}
where all $\omega_{i}>0$ are independent of $\alpha$ and their sum
is equal to 1. In this way $Sp[\rho^{*}(\alpha)]=exp(-\alpha)$.
Then in the momentum representation $\alpha=p^{2}/p^{2}_{max}$,
$p_{max}\sim p_{pl}$,where $p_{pl}$ is the Planck momentum. When
present in matrix elements, $exp(-\alpha)$  damps the contribution
of great momenta in a perturbation theory.
\\It is clear that for each of the coordinates $q_{i}$
the exponential ansatz makes a contribution to the
deformed wave function $\psi(\widetilde{\alpha},q)$ the
modulus of which equals $exp(-\alpha_{i}/2)$  and,
obviously, the same contribution to the conjugate
function $\psi^{*}(\widetilde{\alpha},q)$. Because of
this, for exponential ansatz one may write
\begin{equation}\label{U27S}
\psi(\widetilde{\alpha},q)=\theta(\widetilde{\alpha})\psi(q),
\end{equation}
where $|\theta(\widetilde{\alpha})|=exp(-\sum_{i}\alpha_{i}/2)$.
As noted above, the last exponent of the momentum representation
reads $exp(- \sum_{i}p_{i}^{2}/2p_{max}^{2})$ and in this way it
removes UV (ultra-violet) divergences in the theory. It follows
that $\widetilde{\alpha}$ is a new small parameter. Among its
obvious advantages one could name:
\\1)  its dimensionless nature,
\\2)  its variability over the finite interval
$0<\alpha_{i}\leq 1/4$. Besides, for the well-known physics
it is actually very small: $\alpha\sim 10^{-66+2n}$, where
$10^{-n}$ is the measuring scale.
Here the Planck scale $\sim 10^{-33}cm$ is assumed;
\\3)and finally the calculation of this parameter
involves all three fundamental constants, since by Definition 1 of
section 2 $\alpha_{i}= l_{min}^{2}/x_{i}^{2}$, where $x_{i}$ is
the measuring scale on i-coordinate and $l_{min}^{2}\sim
l_{p}^{2}=G\hbar/c^{3}$.
\\ Therefore, series expansion in $\alpha_{i}$ may be
of great importance. Since all the field components
and hence the Lagrangian will be dependent on
$\widetilde{\alpha}$, i.e.
$\psi=\psi(\widetilde{\alpha}),L=L(\widetilde{\alpha}
)$, quantum
theory may be considered as a theory of lattice
$Lat_{\widetilde{\alpha}}$ and hence of lattice
$Lat^{\tau}_{\widetilde{\alpha}}$.
\section{Deformed Quantum Field Theory and Holographic
Principle}
With the use of this approach for the customary
energies a Quantum Field Theory (QFT) is introduced with a high
degree of accuracy. In our context "customary" means the energies
much lower than the Planck ones.
\\It is important that as the spacing of lattice
$Lat^{\tau}_{\widetilde{\alpha}}$ is decreasing in inverse
proportion to the square of the respective node, for a fairly
large node number $N>N_{0}$  the lattice edge beginning at this
node $\ell_{N,N+1}$ \cite{shalyt6,shalyt7,shalyt9} will be of
length $\ell_{N,N+1}\sim 1/N^{3}$, and by this means edge lengths
of the lattice are rapidly decreasing with the spacing number.
Note that in the large-scale limit this (within any preset
accuracy) leads to parameter $\alpha=0$, pure states and in the
end to QFT. In this way a theory for the above-described lattice
presents a deformation of the originally continuous variant of
this theory as within the developed approach continuity is
accurate to $\approx 10^{-66+2n}$, where $10^{-n}$ is the
measuring scale and the Planck scale $\sim 10^{- 33}cm$ is
assumed. Whereas the lattice per se
$Lat^{\tau}_{\widetilde{\alpha}}$ may be interpreted as a
deformation of the space continuum with the deformation parameter
equal to the varying edge length
$\ell_{\alpha^{1}_{\tau_{1}},\alpha^{2}_{\tau_{2}}}$, where
$\alpha^{1}_{\tau_{1}}$ è $\alpha^{2}_{\tau_{2}}$ are two adjacent
points of the lattice $Lat^{\tau}_{\widetilde{\alpha}}$.
Proceeding from this, all well-known theories including
$\varphi^{4}$, QED, QCD and so on may be studied based on the
above-described lattice in QFT.
\\However, now we need the sublattice
$Lat_{\widetilde{\alpha}}$ in
the momentum representation that may be effectively
used  in our further
reasoning: we operate with the wave functional space in
the momentum representation
$\psi(\widetilde{\bf\alpha_{k}},{\bf k})$, where ${\bf
k}=(k_{x},k_{y},k_{z})$,$\widetilde{\bf\alpha_{k}}=(\alpha
_{x},\alpha_{y},\alpha_{z})$, and
$\alpha_{x}=k_{x}^{2}/p_{pl}^{2}$,$\alpha_{y}=k_{y}^{2}
/p_{pl}^{2}$, $\alpha_{z}=k_{z}^{2}/p_{pl}^{2}$. Here
it is implied that $l_{min}=l_{p}$ and hence
$p_{max}=p_{pl}$. Besides, due to the discrete varying
of the coordinates by steps $2l_{min}=2l_{p}$, the
momenta $k_{i}\sim 1/q_{i}$, where $i=x,y,z$,  are also
discretely varying with the nonunifrom steps $1/l_{N}-
1/(l_{N}+2l_{p})$. In this way we arrive to a Quantum
Field Theory at a nonuniform lattice for all
variables  $(\widetilde{\bf\alpha_{k}},{\bf k})$.
\\ All the variables associated with the considered $\alpha$ -
deformed quantum field theory at the lattice
$Lat_{\widetilde{\alpha}}$ are hereinafter denoted by the upper
index $^{\alpha}$, (i.e.
$QFT^{\alpha}$ is compared to the well-known QFT).
\\Now we show the conformity of $QFT^{\alpha}$ to the holographic
principle, i.e. to the holographic entropy bound  derived in the
earlier works \cite{Hooft1},\cite{Hsu}. As follows from the
holographic principle, the maximum entropy that can be stored
within a bounded region $\Re$ in $3$-space must be proportional to
the value $A(\Re)^{3/4}$, where $A(\Re)$ is the surface area of
$\Re$. Of course, this is associated with the case when the region
$\Re$ is not an inner part of a particular black hole. Provided a
physical system contained in $\Re$ is not bounded by the condition
of stability to the gravitational collapse, i.e. this system is
simply non-constrained gravitationally, then according to the
conventional QFT $S_{\max}(\Re)\sim V(\Re)$, where $V(\Re)$ is the
bulk of $\Re$. As we are considering a lattice, whose coordinates
$(\widetilde{\bf\alpha_{k}},{\bf k})$ comply with
three-dimensional momenta ${\bf k}=(k_{x},k_{y},k_{z})$, any state
$|\Psi \rangle^{\alpha}$ in the Fock space $H^{\alpha}_{F}(\Re)$
of $QFT^{\alpha}$ conforms to the state $|\Psi \rangle$ from the
Fock space of the conventional QFT $H_{F}(\Re)$ : $|\Psi
\rangle^{\alpha}\longleftrightarrow |\Psi \rangle$ or
$\psi(\widetilde{\bf\alpha_{k}},{\bf k})\longleftrightarrow
\psi({\bf k})$, where (for example, \cite{Yurt1})
\begin{equation}\label{Hol2}
|\Psi \rangle = | n_1 ,\; n_2 , \; \cdots \; , \; n_i ,
\; \cdots \; , n_N \rangle \; \; , \; \; \; \; \; \;
n_j \in  N \;
\end{equation}
denoting a state with $n_i$ particles in the
mode $i$. As a consequence, $QFT^{\alpha}$ is actually
determined at the lattice in the momentum space and hence
$H^{\alpha}_{F}(\Re)\subseteq H_{F}(\Re)$. Therefore,
\begin{equation}\label{Hol3}
dim H^{\alpha}_{F}(\Re)\leq dim H_{F}(\Re)
\end{equation}
and in the large-scale limit $\alpha\rightarrow 0$
\begin{equation}\label{Hol4}
\lim\limits_{\alpha\rightarrow 0}dim
H^{\alpha}_{F}(\Re)= dim H_{F}(\Re).
\end{equation}
Taking into account that a maximum entropy is proportional
to the dimensionality logarithm of the Fock space, we have
\begin{eqnarray}\label{Hol5}
S^{\alpha}_{\max} = - k_B {\rm Tr} (\rho_{\max} \log
\rho_{\max}) = k_B \log \dim H^{\alpha}_F (\Re) \leq
\nonumber \\ S_{\max}=k_B {\rm Tr} (\rho_{\max} \log
\rho_{\max} ) = k_B \log \dim H_F (\Re).
\end{eqnarray}
Let us consider the ansatz (7) from \cite{Hsu} that is a subspace
only of those Fock states $|\Psi\rangle$, for which
\begin{equation}\label{Hol6}
\langle \Psi | H | \Psi \rangle  < R,
\end{equation}
where $R$ - characteristic linear size of the system. And in the
energy representation
\begin{equation}\label{Hol6new}
\langle \Psi | H | \Psi \rangle  < \; E_{\max} \; ,
\end{equation}
where $E_{\max} \sim (c^4/G) R$ is an upper bound on the energy,
ensuring that the field $\phi$ is in a stable configuration
against the gravitational collapse to meet the semiclassical
Einstein equations, and $H$ is the Hamiltonian. It is clear that
(\ref{Hol6}) and (\ref{Hol6new}) mean the same considering
$c=G=1$. It should be noted that the spaces $H_{F}(\Re)$ in QFT
and $H^{\alpha}_{F}(\Re)$ in $QFT^{\alpha}$ possess practically
the same set of eigenstates with respect to $H$. But in the second
case this set involves the factors depending on $\alpha$
(formula(\ref{U27S})). In consequence, the condition of
(\ref{Hol6}) is met only by those
$\psi(\widetilde{\bf\alpha_{k}},{\bf k})\in H^{\alpha}_{F}(\Re)$,
for which $\psi(\widetilde{\bf k})$ meets the same condition in
$H_{F}(\Re)$. It seems that there is a single barrier due to the
presence of the exponential factor in the $\alpha$ - deformed
variant (\ref{Hol6}):
\begin{equation}\label{Hol7}
^{\alpha}\langle \Psi | H | \Psi \rangle^{\alpha} =
exp(-{\bf\alpha}) \langle \Psi | H | \Psi \rangle.
\end{equation}
But since we consider a semiclassical approximation, where
$\alpha\rightarrow 0$, all the arguments \cite{Hooft1},\cite{Hsu}
in this case are valid.
\\When we denote in terms of $H^{grav}_{F}(\Re)\subset
H_{F}(\Re)$ the subspace conforming to (\ref{Hol6}), at least in a
semiclassical approximation (i.e. for $\alpha\rightarrow 0$) and
with the use of (\ref{Hol4}) and (\ref{Hol7}), we obtain:
\begin{equation}\label{Hol8}
\lim\limits_{\alpha\rightarrow 0}dim H^{\alpha,grav}_{F}(\Re)= dim
H^{grav}_{F}(\Re).
\end{equation}
Then, as it has been demonstrated in \cite{Hooft1},\cite{Hsu},
\begin{equation}\label{Hol9}
S_{\max}(\Re) \sim \frac{A(\Re)^{3/4}}{{\l_p}^2},
\end{equation}
\\It should be noted that
\\(1)as $QFT^{\alpha}$ at the lattice $Lat_{\widetilde{\alpha}}$
is ultraviolet-finite just from the start there is no need in
regularization of the complete Hamiltonian of the theory $H$;
\\(2) of course, we neglect a
small Casimir-effect contribution to the vacuum stress- energy.
Besides, it is assumed that some general conditions of conformity
to the holographic principle ,e.g. indicated in
\cite{Bou1}--\cite{Wald} are met;
\\(3) one can make the calculations more intricate by the
inclusion of the  $\alpha$-deformed Hamiltonian, similar to
\cite{shalyt7},\cite{shalyt9}. However, it is clearly seen that in
a semiclassical approximation the principal result remains
invariable, as the right-hand side (\ref{Hol7})may be changed only
after the inclusion of some  $\alpha$ - exponents amounting to 1
in the limit $\alpha\rightarrow 0$;
\\(4) the principal objective of this section is to demonstrate
that with the introduction of the fundamental length into a
quantum field theory such a theory may be conforming to the
holographic entropy bound like the conventional QFT
\cite{Hooft1},\cite{Hsu}. This is significant, considering the
prevailing opinion that a future quantum gravitation theory should
involve the holographic principle without fail. But now it is
obvious that such a theory is impossible without the notion of a
fundamental length \cite{Gar1}. In this work we have considered
only a single variant of possible quantum theories with the
fundamental length, that fortunately has proved to be conforming
to the holographic principle with due regard  for to the
gravitational field.
\section{Dark Energy and New Small Parameters}
In terms of the deformation parameter $\alpha$ the principal values
of the Dark Energy Problem may be simply and clearly defined.
Let us begin with the Schwarzschild black holes, whose
 semiclassical entropy is given by
\begin{equation}\label{D1}
S = \pi {R_{Sch}^2}/ l_p^2=\pi {R_{Sch}^2}
M_p^2=\pi\alpha_{R_{Sch}}^{-1},
\end{equation}
with the assumption that in the formula for $\alpha$ $R_{Sch}=x$ is
the measuring scale and $l_p = 1/M_p$. Here $R_{Sch}$
is the adequate Schwarzschild radius, and $\alpha_{R_{Sch}}$
is the value of $\alpha$ associated with this radius. Then, as it
has been pointed out in \cite{Bal}), in case the Fischler-
Susskind cosmic holographic conjecture \cite{Sussk1} is valid,
the entropy of the Universe is limited by its "surface"
measured in Planck units \cite{Bal}:
\begin{equation}\label{D2}
S \leq \frac{A}{4} M_P^2,
\end{equation}
where the surface area $A = 4\pi R^2$ is defined in
terms of the apparent (Hubble) horizon
\begin{equation}\label{D3}
R = \frac{1}{\sqrt{H^2+k/a^2}},
\end{equation}
with curvature $k$  and scale $a$ factors.
\\ Again, interpreting $R$ from (\ref{D3}) as a measuring scale,
we directly obtain(\ref{D2}) in terms of $\alpha$:
\begin{equation}\label{D4}
S \leq \pi\alpha_{R}^{-1},
\end{equation}
where $\alpha_{R}=l^{2}_{p}/R^{2}$. Therefore, the average entropy
density may be found as
\begin{equation}\label{D5}
\frac{S}{V}\leq \frac{\pi \alpha_{R}^{-1}}{V}.
\end{equation}
Using further the reasoning line of \cite{Bal} based on
the results of the  holographic thermodynamics, we can relate
the entropy and energy of a
holographic system \cite{Jac1,Cai1}. Similarly, in terms of the
$\alpha$ parameter one can easily estimate the upper limit for
the energy density of the Universe (denoted here by $\rho_{hol}$):
\begin{equation}\label{D6}
\rho_{hol} \leq \frac{3}{8 \pi R^2} M_P^2 = \frac{3}{8
\pi}\alpha_{R} M_P^4,
\end{equation}
that is drastically differing from the one obtained with a naive
QFT
\begin{equation}\label{D7}
\rho^{naive}_{QFT}\sim M_P^4.
\end{equation}
Here by $\rho^{naive}_{QFT}$ we denote the energy
Density calculated from the naive QFT \cite{Wein1,Zel1}.
Obviously, as $\alpha_{R}$ for $R$ determined by
(\ref{D3}) is very small, actually approximating zero, $\rho_{hol}$
is by several orders of magnitude smaller than the value
expected in QFT - $\rho^{naive}_{QFT}$.
\\In fact, the upper limit of the right-hand side of(\ref{D6})
is attainable, as it has been demonstrated in \cite{Bou3} and
indicated in \cite{Bal}.
The "overestimation" value of $r$ for the energy
density $\rho^{naive}_{QFT}$, compared to
$\rho_{hol}$, may be determined as
\begin{equation}\label{D8}
r =\frac{\rho^{naive}_{QFT}}{\rho_{hol}}=\frac{8
\pi}{3}{\bf \alpha_{R}^{-1}}
 =\frac{8 \pi}{3} \frac{R^2}{L_P^2}
 =\frac{8 \pi}{3} \frac{S}{S_P},
\end{equation}
where $S_P$ is the entropy of the Plank mass and length
for the Schwarzschild black hole. It is clear that due to
smallness of $\alpha_{R}$ the value of $\alpha_{R}^{-1}$
is on the contrary too large. It may be easily calculated
(e.g., see \cite{Bal})
\begin{equation}\label{D9}
r = 5.44\times 10^{122}
\end{equation}
in a good agreement with the astrophysical data.
\\ Naturally, on the assumption that the vacuum energy density
$\rho_{vac}$ is involved in $\rho$ as a term
\begin{equation}\label{vac1}
\rho = \rho_M + \rho_{vac},
\end{equation}
where $\rho_M$ - average matter  density, in case of $\rho_{vac}$
we can arrive to the same upper limit (right-hand side of the
formula(\ref{D6})) as for $\rho$.
\\{\bf Note} Based on the results of \cite{Hooft1},\cite{Hsu},
the entropy estimate in the Fischler-Susskind cosmic holographic
conjecture \cite{Sussk1} is excessive as the equality sign in
(\ref{D2}) may appear in case of the black hole only. And in all
other cases the right-hand side of (\ref{D2}) should contain
$A^{3/4}$ with some factor rather than $A$. It seems that the
entropy estimate obtained in \cite{Sussk1} requires more exact
definition.
\begin{center}
{\bf Discussion}
\end{center}
This section is devoted to the demonstration of the fact, that in
case of the holographic principle validity in terms of the new
deformation parameter $\alpha$ in $QFT^{\alpha}$, considered in
the previous sections and introduced by the author in his works
written as early as 2002
\cite{shalyt11}--\cite{shalyt13},\cite{shalyt1}--\cite{shalyt10},
all the principal values associated with the Dark Energy Problem
may be defined simply and naturally. At the same time, there is no
place for such a parameter in the well-known QFT, whereas in QFT
with the fundamental length, specifically  in $QFT^{\alpha}$ it is
quite natural
\cite{shalyt1,shalyt2,shalyt4,shalyt6,shalyt7,shalyt9}.
\\As indicated in the previous section, $QFT^{\alpha}$ (similar to
the conventional QFT) conforms to the Holographic Principle, being
coincident with QFT to a high accuracy in a semiclassical
approximation, i.e. for $\alpha\rightarrow 0$. In this case
$\alpha$ is small rather than vanishing. Specifically, the
smallness of $\alpha_{R}$ results in a very great value of $r$ in
(\ref{D8}),(\ref{D9}). Besides, from (\ref{D8}) it follows that
there exists some minimal entropy $S_{min} \sim S_P$, and this is
possible  only in QFT with the fundamental length.
\\
\\ It should be noted that this section is related to Section 3
in \cite{Padm1} as well as to Sections 3 and 6 in \cite{Padm2}.
The constant $L_\Lambda$ introduced in these works is such that in
case under consideration $\Lambda\equiv l_\Lambda^{-2}$ is
equivalent to $R$, i.e. $\alpha_{R}\approx \alpha_{l_\Lambda}$
with $\alpha_{l_\Lambda}=l^{2}_{p} /l^{2}_\Lambda$. Then
expression in the right-hand side of (\ref{D6}) is the major term
of the formula for $\rho_{vac}$, and its quantum corrections are
nothing else as a series expansion in terms of
$\alpha_{l_\Lambda}$ (or $\alpha_{R}$):
\begin{equation}\label{D10}
\rho_{\rm vac}\sim
{\frac{1}{l_P^4}\left(\frac{l_P}{L_\Lambda}\right)^2}
+{\frac{1}{l_P^4}\left(\frac{l_P}{l_\Lambda}\right)^4} + \cdots
=\alpha_{l_\Lambda} M_P^4+\alpha^{2}_{l_\Lambda} M_P^4+...
\end{equation}
 In the first variant presented in \cite{Padm1} and
\cite{Padm2} the right-hand side (\ref{D10}) (formulas (12),(33)
in \cite{Padm1} and \cite{Padm2}, respectively)reveals an enormous
additional term $M_P^4\sim \rho^{naive}_{QFT}$ for
renormalization. However, by the approach outlined in Section 5 of
\cite{Padm2} and \cite{Padm3} it may be ignored because the
gravity is described by a pure surface term. And in case under
study, owing to the Holographic Principle, we may proceed directly
to (\ref{D10}). Moreover, in $QFT^{\alpha}$ there is no need in
renormalization as from the start we are concerned with the
ultraviolet-finiteness.
\\ As regards the last part of this section,
(\ref{D10}) in particular, it is expedient to make the following
remark:
\\ in the earlier works by the author
\cite{shalyt6,shalyt7,shalyt9,shalyt10} it was demonstrated that
finding of the quantum correction factors for the primary
deformation parameter $\widetilde{\alpha}$ is a power series
expansion in each $\alpha_{i}$. In the simplest case (Definition
$1^{\prime}$) this means expansion of the left side in the
relation $|\theta(\alpha_{i})|^{2}-|\theta(\alpha_{i})|^{4}\approx
\alpha_{i}$:
\\
$$|\theta(\alpha_{i})|^{2}-
|\theta(\alpha_{i})|^{4}=\alpha_{i}+a_{0}\alpha^{2}_{i}+a_{1}\alpha^{3}_{i}+...$$
\\ and calculation of the associated coefficients
$a_{0},a_{1},...$. The proposed approach to calculation of the
quantum correction factors may be used in the formalism for the
density pro- matrix (Definition 1). In this case the primary
relation \ref{U1} may be written in the form of a series
\begin{equation}\label{U1b} Sp[\rho(\alpha)]-
Sp^{2}[\rho(\alpha)]=\alpha+a_{0}\alpha^{2}
+a_{1}\alpha^{3}+...
\end{equation}
As a result, the measurement procedure using the
exponential ansatz (\ref{U26S}) may be understood as
calculation of the factors $a_{0}$,$a_{1}$,... or
definition of additional members in the exponent
"destroying" $a_{0}$,$a_{1}$,... \cite{shalyt10}. It is
easy to check that the exponential ansatz gives $a_{0}=-
3/2$, being coincident with the logarithmic correction
factor for the Black Hole entropy \cite{r24},\cite{shalyt10}.
Such an approach to calculations of the Quantum Black Hole Entropy
was proposed in the earlier work by the author \cite{shalyt10}.
It is of particular interest that in other work devoted to
calculation of the quantum corrections to the Black Hole
entropy with the help of the Generalized Uncertainty
Principle (GUP)\cite{Med} the parameter $\alpha$
was also, but implicitly, involved.
\section{Some Comments and Additions}
Proceeding from all the above, it is clear that the cosmological
Constant $\Lambda$ should set up the infra-red limit
$R=l_\Lambda=l_{IR}=\sqrt{1/\Lambda}$. Here it is implicitly
stated that $\Lambda$, though being very small, is nonzero.
However, strictly speaking, this fact is not established (e.g.,
see \cite{Bal}). As indicated in the previous works by the author
\cite{shalyt6,shalyt7,shalyt9}, $QFT^{\alpha}$ is
ultraviolet-sensitive rather than infrared-sensitive. Indeed, the
definition of the $\alpha$-deformation involves the minimal length
$l_{min}\sim l_{p}$ and the associated minimal measurable length
$l^{meas}_{min}=2l_{min}$,with the omission of the maximal length
$l_{max}$, i.e. we have $\alpha_{max}=\alpha_{UV}=1/4$, and there
is no $\alpha_{min}=\alpha_{IR}$. Here, as usual, UV and IR denote
ultraviolet and infrared, respectively. Thus, a certain further
modification is required in $QFT^{\alpha}$ to include
$\alpha_{min}\equiv\alpha_{IR}$ in a natural way.  In other words,
it is necessary to have the bound, specified for the lattice
$Lat_{\widetilde{\alpha}}$ by the equation $\alpha=\alpha_{IR}$
($\alpha=\alpha_{R}=\alpha_{l_\Lambda}$ in the considered case),
in terms \cite{shalyt6,shalyt9} of the separated (critical)one.
\\ To this end, of great importance may be the results associated with the explicit description of the stabilized Poincare-Heisenberg
algebra and their application to the cosmological constant
\cite{Ahl2}. This work, based on the results of
\cite{Crysl},\cite{Men},\cite{Crys2}, presents an explicit form of
the  Poincare-Heisenberg algebra extension, stable to
infinitesimal perturbations of the structure constants, to where
$R=l_\Lambda=l_{IR}=\sqrt{1/\Lambda}$ belongs just from the start.
In \cite{Ahl2} this value is denoted by $l_C$.
\\ Also, of particular importance may be the results obtained in the
last few years on the infrared gravity modification (e.g.,\cite{Dvali1},\cite{Rub1},\cite{Nun1},etc.).
\section{Conclusion}
In conclusion it might be well to enumerate the principal findings:
 \\(1) the deformed quantum theory $QFT^{\alpha}$ introduced by the author
in his earlier works
\cite{shalyt1,shalyt2,shalyt4,shalyt6,shalyt7,shalyt9} with the
fulfilled gravitational stability condition (\ref{Hol6}) conforms
to the Holographic Principle (i.e holographic entropy bound
\cite{Hooft1},\cite{Hsu}) , similar to the Local Quantum Field
theory (QFT);
\\
\\(2) the deformation parameter $\alpha$ naturally involved in the
statement of $QFT^{\alpha}$ is one of the principal values, in
terms of which the Vacuum Energy Density Problem $\rho_{\rm vac}$
(Dark Energy) is formulated. Specifically, the smallness of
$\alpha$ at low energies may be used to explained the smallness of
$\rho_{\rm vac}$ as well.
\\
\\The following problems necessitate further studies.
\\I. Further extension of $QFT^{\alpha}$ due to the inclusion of
new parameters or constants: $QFT^{\alpha,...}$.
By author's opinion, a minimum expansion should include the
infrared bound $\alpha_{IR}=\alpha_{R}$. However, in the general
case in $QFT^{\alpha,...}$ some other constants from \cite{Ahl2},
are liable to appear, specifically, $\gamma$ and $l_{u}\equiv\gamma l_{p}$.
\\
\\II. Demonstration that $QFT^{\alpha,...}$ conforms
to the Holographic Principle in case of the gravitational
stability, and an effort to substantiate the  Fischler-Susskind
cosmic holographic conjecture \cite{Sussk1} or its refinement (note
in section 4).
\\
\\III. With the findings of points II and III, solution of
the following two problems:
\\(a) Why the vacuum energy is nonzero?
\\(b) Why the vacuum and matter energy densities are
about the same today?
\\ In this work it was implicitly understood that $\rho_{\rm vac}\neq 0$. But presently this is open to dispute, as there is still no proof
for this statement. The first problem is actually formulated as
"What is a nonzero lower bound for \\ $\rho_{\rm vac}\neq 0$?" By
now this problem has no answer (e.g., \cite{Hor}).
\\
\\
{\bf Acknowledgement:}I am most grateful to Dr. Andreas Aste
from the Department of Physics and Astronomy Theory Division,
University of Basel (Switzerland) for his kindness and assistance,
especially for the indication of a subtle mistake in the principal
estimate of the holographic entropy  in \cite{Yurt1}
that has been revealed by him and published in \cite{Aste1}.


\end{document}